\documentclass[twoside]{article}
\begin{document}
\setlength{\textheight}{8.0truein}    %FOR 2ND PAGE ONWARDS

%\runninghead{ }
            { }

%\normalsize\textlineskip
\thispagestyle{empty}
\setcounter{page}{1}

%\copyrightheading{Vol.}{No.}{Year}{Page Nos.}
%\copyrightheading{0}{0}{2003}{000--000}

\vspace*{0.88truein}

%\alphfootnote

%\fpage{1}

\centerline{\bf
%%%%%%%%%%%%%%%%%%%%%
%Put in titiles here
%%%%%%%%%%%%%%%%%%%%%
Getting something out of nothing}
%\vspace*{0.035truein}
%\centerline{\bf FOR QUANTUM INFORMATION AND
%COMPUTATION\footnote{Typeset the title in 10 pt Times Roman,
%uppercase and boldface.}}
\vspace*{0.37truein} \centerline{\footnotesize
%%%%%%%%%%%%%%%%%%%%%%%%%%%%%%%%%%%%
%put authors' name and address here
%%%%%%%%%%%%%%%%%%%%%%%%%%%%%%%%%%%%
Hoi-Kwong Lo \footnote{ Email: hklo@comm.utoronto.ca  }}
\vspace*{0.015truein} \centerline{\footnotesize\it Center for
Quantum Information and Quantum Control} \baselineskip=10pt
\centerline{\footnotesize\it Dept. of Electrical \& Computer
Engineering \& Dept. of Physics} \baselineskip=10pt
\centerline{\footnotesize\it  University of Toronto, 10 King's
College Road } \baselineskip=10pt \centerline{\footnotesize\it
Toronto, Ontario, CANADA, M5S 3G4.
%\footnote{State completely without abbreviations, the
%affiliation and mailing address, including country. Typeset in 8
%pt Times Italic.}
}

% \vspace*{0.225truein} \publisher{(received date)}{(revised
%date)}

\vspace*{0.21truein}

%% \abstracts{first paragraph}{second paragraph}{third paragraph}
%% If there is only one paragraph, just keep the second and third empty
%% like the following one
\abstract{
%\abstracts{
%%%%%%%%%%%%%%%%%%%%
% put abstract here
%%%%%%%%%%%%%%%%%%%%
We study quantum key distribution
with standard weak coherent states and
show, rather counter-intuitively, that the detection events
originated from vacua can contribute to secure key generation rate,
over and above the best prior art result.
Our proof is based on a communication complexity/quantum memory argument.
}{}{}

\vspace*{10pt}

%\keywords{Quantum Cryptography, Quantum Key Distribution,
%Unconditional Security} \vspace*{3pt} \communicate{to be filled by
%the Editorial}

%\vspace*{1pt}\textlineskip    %) USE THIS MEASUREMENT WHEN THERE IS
   %) A SECTION HEADING
%\vspace*{-0.5pt}
%\noindent
%%%%%%%%%%%%%%%%%%%%%%%%%%%%%%%%
%put the text of the paper here
%%%%%%%%%%%%%%%%%%%%%%%%%%%%%%%%
\section{Introduction}
\label{S:Intro}

The best-known application of quantum cryptography is quantum key
distribution (QKD) \cite{BB84}. The goal of QKD is to allow two parties, Alice
and Bob, to share a common string of secret in the presence of an
eavesdropper, Eve. Such a key can subsequently be used for,
for example, perfectly
secure communications via the so-called one-time-pad.
Unlike conventional cryptography, the security of QKD is guaranteed by the
fundamental law of physics---the Heisenberg uncertainty principle.
The best-known protocol for QKD is the Bennett-Brassard protocol (BB84) \cite{BB84}.
In BB84, Alice sends Bob a sequence of single photons in one of
the four polarizations (vertical, horizontal, 45-degree and 135-degree)
and Bob randomly performs a measurement in one of the two
conjugate bases. In principle, the security of QKD has been
proven in a number of papers including \cite{proofs,Ben-Or,ShorPreskill}.

For practical implementations, an attenuated laser pulse
(a so-called weak coherent state) is often used as the
source. The security of QKD with a rather generic class of
imperfect devices has been proven in GLLP \cite{GLLP}, following the
earlier work \cite{ilm}
Recently, Hwang \cite{HwangDecoy} has proposed a decoy state
idea for improving
the performance (i.e., the key generation rate and distance) of
QKD systems. We \cite{Decoy} have demonstrated rigorously
how the decoy state idea
can be combined with GLLP to obtain a key generation rate
(per pulse emitted by Alice) which
is
lower bounded by:
\begin{equation}\label{refinedkeyrate}
S \geq Q_{signal} \{- H_2(E_{signal}) + \Omega_1 [ 1- H_2(e_1)] \},
\end{equation}
where $Q_{signal}$ and $E_{signal}$ are respectively the gain and
quantum bit error rate (QBER) of the signal state, $\Omega_1$ and
$e_1$ are respectively the fraction and QBER of detection events by
Bob that have originated from single-photon signals emitted by
Alice. Here, the gain means the ratio of
Bob's detection events to Alice's total number of emitted signals.
[Decoy state QKD has subsequently been investigated by
Wang \cite{WangDecoy} and by Harrington \cite{Harrington}.]

The key goal of this paper is to increase the above key generation
rate in Eq.~\ref{refinedkeyrate} by a term $Q_{signal} \Omega_0$
where $\Omega_0$ is the fraction of detection events of Bob that
have originated from vacua emitted by Alice. More concretely,
we have the following main Theorem.

{\bf Theorem~1} The key generation of an efficient BB84 scheme is
given by:
\begin{equation}\label{newkeyrate}
S \geq Q_{signal} \{- H_2(E_{signal}) + \Omega_0 + \Omega_1 [ 1- H_2(e_1)] \},
\end{equation}
where $\Omega_0$ is the fraction of detection events by Bob that
has originated from the vacuum signals emitted by Alice.
In other words, we find that each detection event by Bob that has
originated from a
vacuum (i.e., nothing) emitted by Alice automatically contributes to a bit of
secure key over and above the prior art result
(Eq.~\ref{refinedkeyrate})
presented in \cite{GLLP}
and also \cite{Decoy}.

Before we embark on a detailed discussion, let us check for
the consistency of our result. Naively, one might think that
our suggestion that the vacuum will contribute to a secure
key is an insane idea because if nothing is emitted by Alice,
what is the origin of security? We remark that the vacuum {\it alone}
does {\it not} contribute to a secure key. More concretely, suppose
all the signals sent by Alice are vacua and there are no background
events. Then, $\Omega_0 =1$, $\Omega_1 =0$, and $E_{signal} = 1/2$.
Therefore, from Eq.~\ref{newkeyrate}, we get the lower bound $0$ for
the key generation rate. The reason is that the term $\Omega_0$ is
exactly cancelled by the error correction term $- H_2(E_{signal})$.

What Eq.~\ref{newkeyrate} does say is that no privacy amplification
is needed for the vacua state. This is intuitively clear because
Eve cannot have any a priori information on Alice's bit, if nothing is
emitted from Alice's laboratory.

Now, let us prove our main result (Eq.~\ref{newkeyrate}).
We shall use the method of
communication complexity.
As noted by by Ben-Or \cite{Ben-Or} and
by Renner and Koenig \cite{RennerKoenig}, the number of
rounds of universal hashing needed for privacy amplification
in QKD is at most given by any upper bound to the size of
Eve's quantum memory which contains information
on the key. In other words, we have, informally:

{\bf Theorem~2} \cite{Ben-Or,RennerKoenig}: The key generation rate in QKD
\begin{equation}
S \geq  N-{\cal S}_{Eve}
\end{equation}
where $N$ is the size of the sifted key shared between Alice
and Bob and ${\cal S}_{Eve}$ is the size of Eve's quantum memory.
[A more formal definition involving the relevant $\epsilon$ and
$\delta$ can be found as Eq.~(11) in \cite{RennerKoenig}.]

{\it Remark}: Note that Theorem~2 only gives a lower bound to the key
generation rate because it does not consider the possibility of
advantage distillation in QKD \cite{twoway}.

In summary, all we need to compute (a lower bound to) the
key generation rate is to work out the size of Eve's
quantum memory.

{\bf Proof of Theorem~1}:
Now, note that Eve has two pieces of information on the key.
The first piece, which is strictly quantum,
comes from Eve's eavesdropping attack
during the quantum transmission from Alice to Bob.
The second piece is classical and comes from the classical
error correction part.

We argue that the first piece, from eavesdropping the
quantum transmission, consists of
two parts: single-photon part and multi-photon part.
It should be emphasized that the vacua signals do {\it not} contribute
at all. This is because, since Alice is emitting nothing,
Eve cannot possibly learn anything about Alice's key.
Eve can influence and, in fact, decide on Bob's key by
sending her own photons into Bob's detector. However,
Bob's key does not really tell Eve anything about Alice's key.

Let us consider the multi-photon part first.
We take the most conservative assumption that Eve has all the
information on all multi-photon signals. Her quantum memory size on
the multi-photon part is then given by
$Q_{signal} \Omega_m$ .
Here, $\Omega_m$ is the fraction of detection events of
Bob that have originated from multi-photon signals.
Note that $\Omega_0 + \Omega_1 + \Omega_m = 1$.
The single-photon part is given by simply
$ Q_{signal} \Omega_1 H_2 (e_1^{phase})$, where
$e_1^{phase}$ is the phase error rate of the single-photon
signals. From Shor-Preskill's proof \cite{ShorPreskill},
$e_1^{phase} = e_1$, which is the bit-flip error rate for
the single-photon signals. So, the quantum memory for
single-photon part is actually given by
$ Q_{signal} \Omega_1 H_2 (e_1)$.
Adding the two parts, the first piece of Eve's information has a
memory size $Q_{signal} [ \Omega_1 H_2 (e_1) + \Omega_m ]$.
The second piece of Eve's information, which comes from classical
error correction, is asymptotically
given by $Q_{signal} H_2(E_{signal})$.

In summary,
adding the two pieces together, the total quantum memory size of Eve
is given by ${\cal S}_{Eve}=
Q_{signal} [ H_2(E_{signal}) + \Omega_1 H_2 (e_1) + \Omega_m ]$.

Now, the length of the sifted key (per pulse emitted by Alice)
shared by Alice and Bob is
$N= Q_{signal} [ \Omega_0 + \Omega_1 + \Omega_m ]$.
Therefore, the number of secure key bits (per pulse emitted
by Alice) is given by
\begin{eqnarray}
S &\geq&  N-{\cal S}_{Eve} \nonumber \\
  &=& Q_{signal} \{ - H_2(E_{signal}) + \Omega_0 + \Omega_1 [ 1- H_2(e_1)] \}.
\end{eqnarray}
which is precisely Eq.~(\ref{newkeyrate}).
This concludes the proof of our Theorem~1.

In summary, we have increased the key generation from
Eq.~(\ref{refinedkeyrate}) in
the prior art result
 \cite{GLLP} to Eq.~(\ref{newkeyrate}) by showing that, rather
 counter-intuitively, the detection
events due to vacua contribute directly to the secure key.
What is interesting about this result is that it is based on a
communication complexity approach and is not entirely clear whether it
can be derived from an entanglement distillation approach.
In future, it will perhaps be interesting to rephrase this result in the
general framework of $\Gamma$ states \cite{Gammastates}, which
generalizes the entanglement distillation approach.

%\nonumsection
\section*{Acknowledgements} \noindent We thank helpful
discussions with colleagues including J. Batuwantudawe,
Jean-Christian Boileau, Debbie Leung,
John Preskill and Kiyoshi Tamaki.
This part is financially supported in part
by funding agencies including CFI, CIPI, CRC
program, NSERC, OIT, and PREA.
Parts of this paper were written during visits to the Institute of
Quantum Information (IQI) at Caltech
and to the Isaac Newton Institute, Cambridge, UK, whose kind hospitality is
acknowledged.

\end{document}